\documentstyle[graphicx]{mn}

\newcommand{\newblock}{}

\newcommand\aproxgt{\mathrel{%
      \rlap{\raise 0.511ex \hbox{$>$}}{\lower 0.511ex \hbox{$\sim$}}}}
\newcommand\aproxlt{\mathrel{%
      \rlap{\raise 0.511ex \hbox{$<$}}{\lower 0.511ex \hbox{$\sim$}}}}
\def\errtwo#1#2#3{${#1}^{+#2}_{-#3}$}

\newcommand\gx{GX~339$-$4}
\newcommand\cyg{Cyg~X-1}

\newcommand\rxte{\textsl{RXTE}}

\begin{document}

\title{Are There Three Peaks in the Power Spectra of \gx\ and \cyg?}
\author[M.A. Nowak]{
  M.A.~Nowak$^1$ \\
  $^1$ JILA, University of  Colorado, Campus Box 440, Boulder, CO
  80309-0440, U.S.A. 
}

\maketitle

\begin{abstract}  
  Among the variability behaviour exhibited by neutron star systems are the
  so-called ``horizontal branch oscillations'' (HBO, with frequencies
  $\approx 50$\,Hz), the ``lower-frequency kHz quasi-periodic oscillation''
  (QPO) and the ``upper-frequency kHz QPO'', with the latter two features
  being separated in frequency by an amount comparable to, but varying
  slightly from, the suspected spin-frequency of the neutron star.
  Recently, Psaltis, Belloni, \& van der Klis (1999) have suggested that
  there exists a correlation between these three frequencies that, when
  certain identifications of variability features are made, even
  encompasses black hole sources.  We consider this hypothesis by
  reanalyzing a set of \gx\ observations. The power spectral density (PSD)
  constructed from a composite of 7 separate, but very similar,
  observations shows evidence for three broad peaks in the PSD.  If the
  peak frequencies of these features are identified with ``QPO'', then
  their frequencies approximately fit the correlations suggested by
  Psaltis, Belloni, \& van der Klis (1999).  We also reanalyze a \cyg\ 
  observation and show that the suggested QPO correlation may also hold,
  but that complications arise when the ``QPOs'' (which, in reality, are
  fairly broad features) are considered as a function of energy band.
  These fits suggest the existence of at least three separate, independent
  physical processes in the accretion flow, a hypothesis that is also
  supported by consideration of the Fourier frequency-dependent time lags
  and coherence function between variability in different energy bands.  If
  these variability features have a common origin in neutron star and black
  hole systems, then ``beat frequency models'' of kHz QPO in neutron star
  systems are called into question.
\end{abstract}


\section{Introduction}\label{sec:intro}

Observations by the \textsl{EXOSAT} and later by the \textsl{Ginga}
satellites during 1980's and early 1990's, and more recently by the
\textsl{Rossi X-ray Timing Explorer} (\rxte) have revealed a rich variety
of X-ray variability behaviour in neutron star and black hole candidate
(BHC) systems.  Neutron star systems in general, and the so-called
Z-sources in specific \cite{hasinger:89a}, have exhibited a set of
relatively narrow features in their X-ray variability power spectral
densities (PSD).  These features, referred to as quasi-periodic
oscillations (QPO), phenomenologically appear to fall predominantly into
one of four classes.  At low Fourier frequencies (see van der Klis
\nocite{vanderklis:95a} 1995 for a review) there are the so-called normal
branch oscillations (NBO) with $f\approx 5$--$20$\,Hz and the horizontal
branch oscillations (HBO) with $f\approx 15$--$60$\,Hz.  A second set of
QPO, typically occurring in pairs at high frequencies $f\approx
200-1200$\,Hz, are referred to as the ``lower-frequency kHz QPO'' and the
``upper-frequency kHz QPO'' \cite{vanderklis:96a}. Similar high frequency
features have been observed in lower-luminosity neutron star systems,
specifically the so-called atoll sources \cite{strohmayer:96a}.  The atoll
systems also have exhibited low-frequency features that in some ways are
similar to the HBO (see Hasinger \& van der Klis \nocite{hasinger:89a}
1989, Homan et al.  \nocite{homan:98a} 1998).

A wide variety of features ranging from narrow (see, for example, Nowak,
Wilms, \& Dove \nocite{nowak:99c} 1999; hereafter NWD) to broad ``noise
components'' (see, for example, the discussion of van der Klis
\nocite{vanderklis:94a} 1994a, van der Klis \nocite{vanderklis:94b} 1994b)
have been observed in BHC systems as well.  Frequently, the PSDs associated
with these systems are flat at low frequencies, show a low frequency break
into a steeper $f^{-1}$--$f^{-2}$ spectrum, and often exhibit a low
frequency QPO at frequencies slightly above the break (see Wijnands \& van
der Klis \nocite{wijnands:99a} 1999, and references therein).  It has been
noted that the location of the break and the frequency of the QPO are often
correlated \cite{wijnands:99a}, both in neutron stars and BHC systems.

In a recent work, Psaltis, Belloni, \& van der Klis \shortcite{psaltis:99a}
(hereafter PBK) have presented a further analogy between the features
observed in neutron star and in BHC systems.  Specifically, they have shown
that the HBO frequency is apparently correlated with the lower-frequency
kHz QPO frequency (Fig. 1 of PBK). With suitable identifications (in part
relying upon the break frequency-QPO frequency correlation of Wijnands \&
van der Klis \shortcite{wijnands:99a}, as discussed by PBK) to features
observed in atoll and BHC systems, this correlation is seen to extend over
three decades in Fourier frequency and encompass Z-sources, atoll sources,
and BHC sources.  If these apparent correlations exist because of an
underlying common physical mechanism, then all these variability phenomena
are somehow intrinsic to the accretion flow, and do not explicitly rely
upon the presence or absence of either a hard surface or a magnetic field
in these systems.

A correlation is also observed between the lower-frequency kHz and
upper-frequency kHz QPO.  Some models (e.g., Miller, Lamb, \& Psaltis
\nocite{miller:98a} 1998) associate this difference frequency with being
nearly equal to the spin-frequency of the neutron star (in contrast to the
theories of Psaltis \& Norman \nocite{psaltis:99b}, Stella \& Vietri
\nocite{stella:98a} 1998, Merloni et al. \nocite{merloni:99aa} 1999, and
Stella et al.  \nocite{stella:99a} 1999 discussed further below).  PBK,
however, speculated about whether the observed lower-frequency kHz
QPO/upper-frequency kHz QPO correlation would also extend to low
frequencies and BHC systems in a manner similar to the putative
HBO/lower-frequency kHz QPO correlation.  In this work we reexamine a set
of observations of the BHC \gx\ in order to search for evidence of a
``third QPO'' in this system.  We present such evidence in \S\ref{sec:psd}.
In \S\ref{sec:psdcyg}, we also present evidence for multiple broad-peaked
features in the PSD of \cyg.  We discuss in \S\ref{sec:discuss} the
significance of these fits as regards the hypothesized QPO correlations.
Furthermore, in light of the observed time lags and degree of linear
correlation (i.e., the coherence function) between the soft and hard X-ray
variability, we discuss whether the individual ``QPO fit-components'' to
the PSD actually represent physically distinct processes in the accretion
flow.  We summarize our results in \S\ref{sec:summary}.

\section{Power Spectral Densities for \gx}\label{sec:psd}

In a previous work (NWD), we presented timing analysis for eight separate
\rxte\ observations of \gx.  The faintest observation, which in terms of
3--9\,keV flux is a factor $\sim 2.5$ fainter than the next brightest
observation, clearly shows larger amplitude and characteristically
lower-frequency variability than the other observations.  Furthermore, the
faintest observation shows time lags between its soft and hard X-ray
variability \cite{miyamoto:89a,miyamoto:92a,vaughan:97a,nowak:99a} that are
significantly shorter than those exhibited by the other observations. The
remaining observations, which span a factor of $\sim 2$ in terms of
3--9\,keV flux, have relatively similar timing properties.  Specifically,
the shapes and amplitudes of their PSDs are all comparable, and each
shows a narrow PSD peak between 0.26--0.34\,Hz.  In addition, their Fourier
frequency-dependent time lags span less than a factor of two at any
given Fourier frequency.

In the frequency range of 0.1--30\,Hz, NWD were relatively successful in
fitting the PSDs to a functional form that consisted of a power law
approximately $\propto f^{-1}$, plus two additional Lorentzian components
of the form
\begin{equation}
P(f) = \pi^{-1} \frac{R^2 Q f_0}{f_0^2 + Q^2 (f - f_0)^2} ~~,
\label{eq:lorentz}
\end{equation}
where $f_0$ is the resonant frequency of the Lorentzian, $Q$ is the quality
factor ($\approx f_0/\Delta f$, where $\Delta f$ is the
full-width-half-maximum of the Lorentzian), and $R$ is the fit amplitude
(root mean square variability, rms $= R [ 1/2 - \tan^{-1}(-Q)/\pi]^{1/2}$,
i.e. rms $=R$ as $Q \rightarrow \infty$).  However, due to signal-to-noise
considerations, NWD were unable to obtain an adequate fit above $\approx
30$\,Hz.

As the seven brightest observations of \gx\ presented in NWD are so
intrinsically similar, we decided to average these observations to form a
composite PSD.  The PSD, utilizing energy channels covering 0--21.9\,keV
(i.e., channels A--D from NWD summed) were calculated and averaged together
using the normalization of Leahy et al. \shortcite{leahy:83a} (note that
this normalization weights the signal PSD by the count rate of the
observation, with the average count rates here ranging from 517--877\,cps),
the noise was subtracted (see Nowak et al. \nocite{nowak:99a} 1999a; and
references therein), and then the PSD was renormalized to the normalization
of Belloni \& Hasinger \shortcite{belloni:90a} using the mean count rate of
the summed observations.  For this normalization, integrating over Fourier
frequency yields the mean square variability relative to the square of the
mean of the lightcurve. Error bars on the PSD were calculated as in Nowak
et al. \shortcite{nowak:99a}

\begin{figure}
\resizebox{\hsize}{!}{\includegraphics{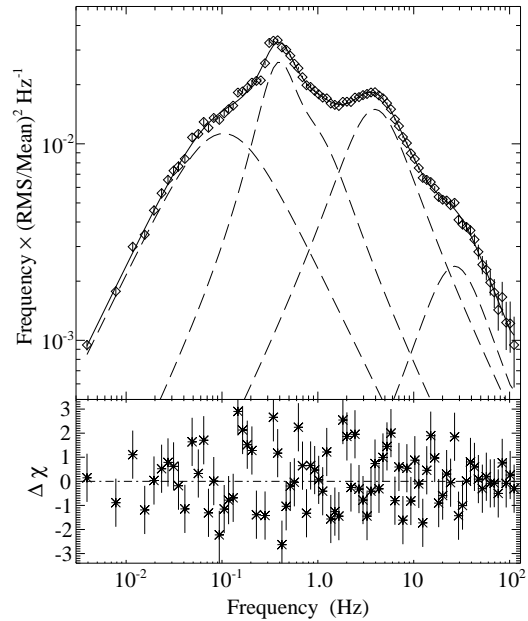}}
\caption{\small Fourier frequency times composite power spectral density
  (PSD) of \gx.  Thick solid line is the best fit model, whereas the dashed
  lines are the separate components of the model. The lower panel shows the
  residuals from the fit.
  \protect{\label{fig:comppsd}}}
\end{figure}

We fit this PSD with a model that consisted of a zero frequency-centered
Lorentzian ($\equiv A_0/[1+(f/f_0)^2]$), plus four additional QPO features
of the form of eq.~\ref{eq:lorentz}.  The two lowest frequency QPO features
were constrained to be harmonics of one another (this allowed us to fit the
``asymmetric'' nature of this feature that was commented upon in NWD), but
otherwise all QPO amplitudes, widths, and frequencies were allowed to be
fit parameters.  Our best fit model is also presented in
Fig.~\ref{fig:comppsd}. The fit yielded $\chi^2=177$ for 75 degrees of
freedom.  The fit parameters (with subscripts 0, 1, h, 2, 3, for the zero
frequency-centered Lorentzian, lowest frequency QPO, its harmonic, and
higher frequency QPOs) are presented in Table~\ref{tab:qpo}.  Error bars
are 90\% confidence level, i.e. $\Delta \chi^2=2.71$ for one interesting
parameter.

\begin{table*}
\caption{\small Parameters for PSD Fits. All frequencies are units of
  Hz. \protect{\label{tab:qpo}}} 
\center{
\begin{tabular}{ccccccc}
\hline
\hline
\noalign{\vskip 3pt}
Observations & zfc-Lorentz. & ${\rm QPO_1}$ & ${\rm QPO_h}$ &
${\rm QPO_2}$ & ${\rm QPO_3}$ & $\chi^2/$DoF \\
\noalign{\vskip 3pt}
\hline
\gx: & $A_0=~$\errtwo{0.22}{0.01}{0.01} &
$R_1=~$\errtwo{0.20}{0.01}{0.01} & $R_h=~$\errtwo{0.14}{0.01}{0.01} &
$R_2=~$\errtwo{0.21}{0.01}{0.01} & $R_3=~$\errtwo{0.08}{0.01}{0.01} &
$177/75$ \\
Composite of & $f_0=~$\errtwo{0.10}{0.01}{0.01} &  $f_1=~$\errtwo{0.34}{0.01}{0.01} & &
$f_2=~$\errtwo{2.5}{0.2}{0.2} & $f_3=~$\errtwo{18}{3}{4} \\
7 Brightest && $Q_1=~$\errtwo{1.7}{0.2}{0.1} & $Q_h=~$\errtwo{0.9}{0.2}{0.1} &
$Q_2=~$\errtwo{0.8}{0.1}{0.1} & $Q_3=~$\errtwo{0.9}{0.3}{0.2} \\
\noalign{\vskip 3pt}
\hline
\noalign{\vskip 3pt}
\gx: & $A_0=~$\errtwo{0.84}{0.18}{0.15} &
$R_1=~$\errtwo{0.19}{0.04}{0.03} & 
$R_h=~$\errtwo{0.16}{0.17}{0.15} & $R_2=~$\errtwo{0.27}{0.02}{0.25} &
$R_3=~$\errtwo{0.04}{0.17}{0.04} & $80/67$ \\
\noalign{\vskip 3pt}
Faintest & $f_0=~$\errtwo{0.03}{0.01}{0.01} &
$f_1=~$\errtwo{0.10}{0.01}{0.02} & & 
$f_2=~$\errtwo{0.9}{3.5}{0.2} & $f_3=~$\errtwo{5}{19}{5} \\
\noalign{\vskip 3pt}
& & $Q_1=~$\errtwo{1.8}{0.8}{0.7} & $Q_h=~$\errtwo{0.8}{0.3}{0.3} &
$Q_2=~$\errtwo{0.5}{1.9}{0.1} & $Q_3=~$\errtwo{0.2}{0.9}{0.2} \\
\noalign{\vskip 3pt}
\hline
\noalign{\vskip 3pt}
\cyg:  & $A_0=~$\errtwo{0.11}{0.00}{0.01} &
$R_1=~$\errtwo{0.02}{0.08}{0.02} & 
$R_h=~$\errtwo{0.09}{0.02}{0.09} & $R_2=~$\errtwo{0.20}{0.06}{0.05} &
$R_3=~$\errtwo{0.06}{0.02}{0.05} & $126/57$ \\
\noalign{\vskip 3pt}
0--4 keV & $f_0=~$\errtwo{0.33}{0.01}{0.04} &
$f_1=~$\errtwo{0.79}{0.06}{0.09} & & 
$f_2=~$\errtwo{0.7}{1.3}{0.6} & $f_3=~$\errtwo{29}{33}{27} \\
\noalign{\vskip 3pt}
& & $Q_1=~$\errtwo{2.6}{10}{2.6} & $Q_h=~$\errtwo{2.3}{3.0}{1.0} &
$Q_2=~$\errtwo{0.3}{0.2}{0.2} & $Q_3=~$\errtwo{0.5}{1.9}{0.5} \\
\noalign{\vskip 3pt}
\hline
\noalign{\vskip 3pt}
Cyg~X-1:  & $A_0=~$\errtwo{0.061}{0.002}{0.002} &
$R_1=~$\errtwo{0.003}{0.011}{0.003} & 
$R_h=~$\errtwo{0.13}{0.01}{0.01} & $R_2=~$\errtwo{0.15}{0.02}{0.02} &
$R_3=~$\errtwo{0.05}{0.01}{0.01} & $154/57$ \\
\noalign{\vskip 3pt}
14--45 keV & $f_0=~$\errtwo{0.36}{0.01}{0.01} &
$f_1=~$\errtwo{0.83}{0.05}{0.05} & & 
$f_2=~$\errtwo{4.2}{1.3}{1.5} & $f_3=~$\errtwo{43}{6}{8} \\
\noalign{\vskip 3pt}
& & $Q_1=~$\errtwo{1.5}{6.2}{1.4} & $Q_h=~$\errtwo{1.3}{0.2}{0.2} &
$Q_2=~$\errtwo{0.6}{0.2}{0.2} & $Q_3=~$\errtwo{2.0}{1.4}{0.8} \\
\noalign{\vskip 3pt}
\hline
\end{tabular}
}
\end{table*}

\begin{figure}
\resizebox{\hsize}{!}{\includegraphics{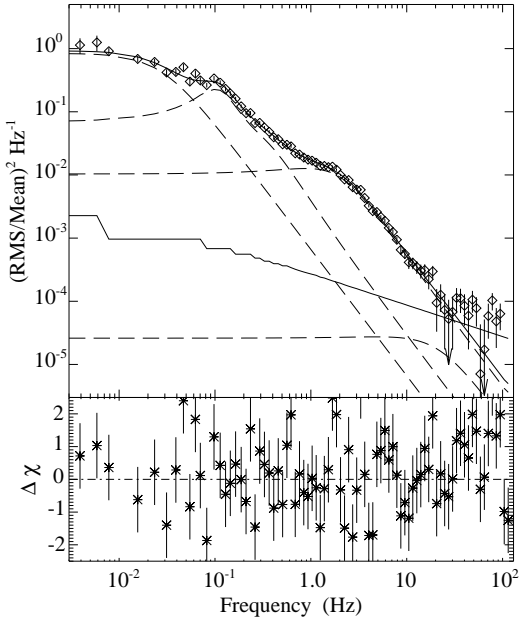}}
\caption{\small Power spectral density (PSD) of \gx\ for the faintest
  observation presented in Nowak et al. \protect{\shortcite{nowak:99c}}.
  Thick solid line is the best fit model, whereas the dashed lines are the
  separate components of the model. The thin solid line is the residual
  noise level (see text) after subtracting the mean noise PSD. Lower panel
  shows the residuals from the fit.  \protect{\label{fig:lowpsd}}}
\end{figure}

We have also applied the above fit to the faintest observation (observation
5) from \nocite{nowak:99c} NWD.  Here we are able to fit the zero
frequency-centered Lorentzian, the low frequency QPO (plus harmonic), and
the middle frequency QPO. The results are also presented in
Table~\ref{tab:qpo}, and the data are shown in Fig.~\ref{fig:lowpsd}.  The
high-frequency QPO, if present, is essentially unconstrained (although in
Table~\ref{tab:qpo} we give the error bars for the ``local minimum'' in
$\chi^2$).  Fig.~\ref{fig:lowpsd} also shows the residual noise level.
This level corresponds to the expected amplitude of postive 1-$\sigma$
fluctuations above the mean value of the Poisson noise PSD, and therefore
is indicative of the minimum PSD amplitude at which a signal can be
detected (see Nowak et al.  \nocite{nowak:99a} 1999a, and references
therein).  This residual noise level, which scales inversely proportionally
to the count rate and to the square root of the integration time, shows the
difficulty of detecting the presence of a ``third QPO feature'' in the high
frequency PSD of a single observation.  The composite PSD shown in
Fig.~\ref{fig:comppsd} has a residual noise level $\approx9$ times lower
than that for the faintest observation of \gx.

\section{Power Spectral Densities for \cyg}\label{sec:psdcyg}

\begin{figure*}
\centerline{
\includegraphics[width=0.33\textwidth]{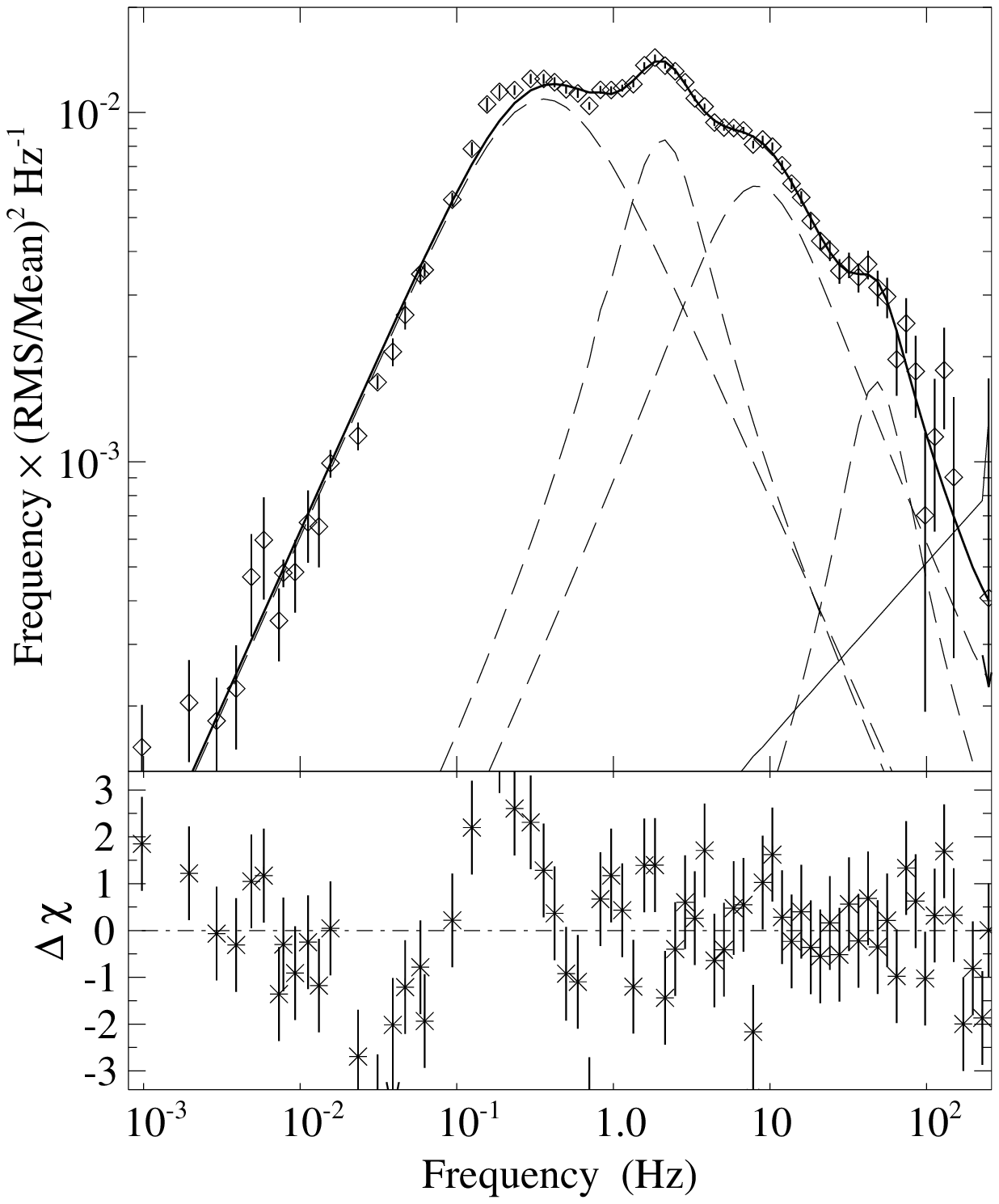}
\includegraphics[width=0.33\textwidth]{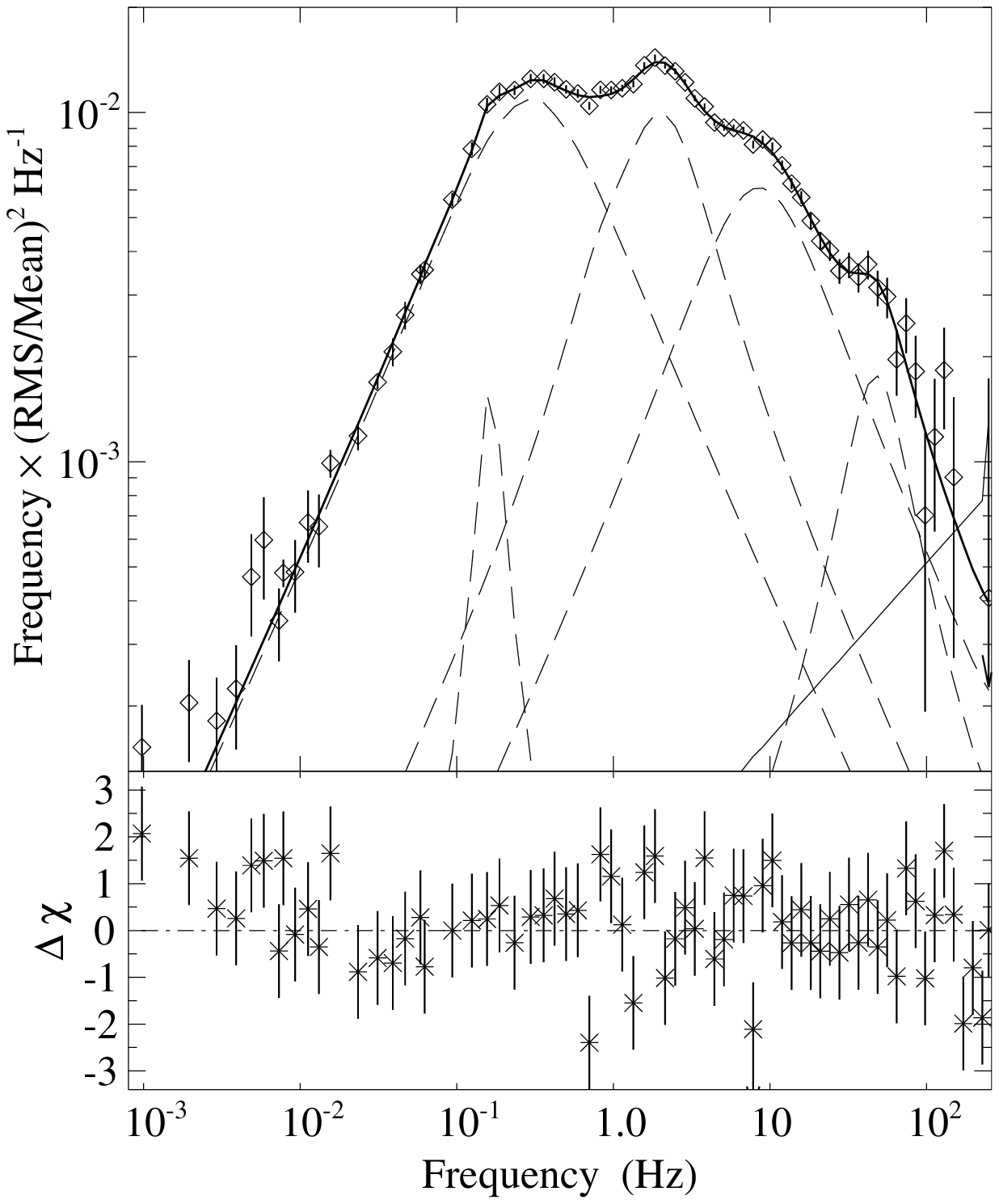}
\includegraphics[width=0.33\textwidth]{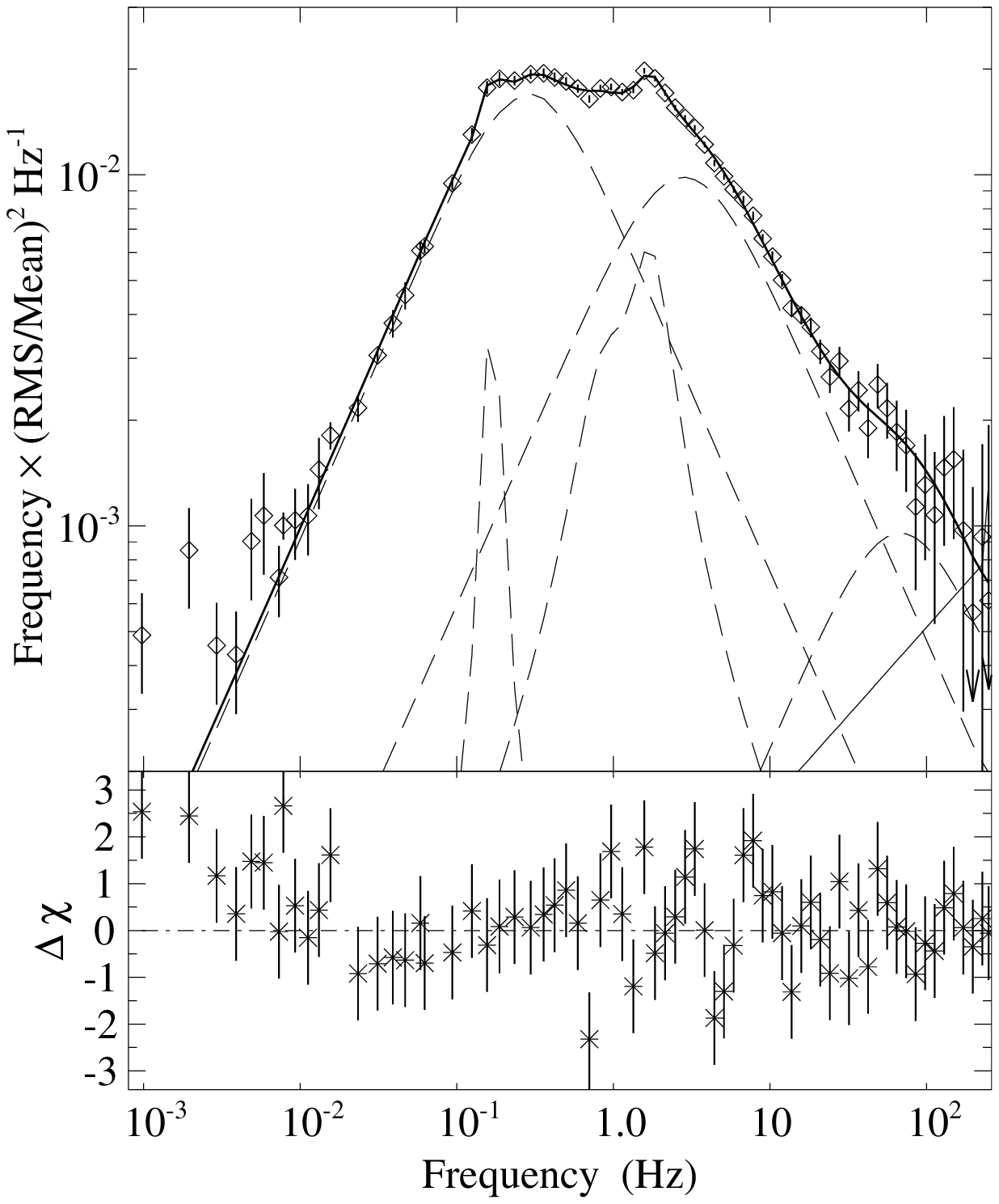}
}
\caption{\small Fourier frequency times the power spectral density (PSD) of
  \cyg\ for the 14--45\,keV band (left, middle) and the 0--4\,keV band
  (right).  Data are the same as presented in Nowak et al. (1999a).  Thick
  solid lines are the best fit models from Table~1 (left) and Table~2
  (middle, right), whereas the dashed lines are the separate components of
  the models. The thin solid lines are the effective noise levels after
  subtracting the mean noise PSDs. Lower panels show the residuals of the
  fits.  \protect{\label{fig:cyg}}}
\end{figure*}

Nowak et al. \shortcite{nowak:99a} discussed observations of \cyg\ in its
low luminosity/spectrally hard state. In that work, we presented doublely
broken power law fits to the PSD, with the low frequency PSD being
essentially flat, the middle-frequency PSD being approximately $\propto
f^{-1}$, and the high-frequency PSD being approximately $\propto f^{-2}$.
Although the \emph{fractional} deviation of the data from the fits was
quite small, the reduced $\chi^2$ ranged from $\approx 4$--$16$ due to the
excellent statistics achievable with RXTE.  Here we refit the PSD with the
functional form discussed in \S\ref{sec:psd}. The fit parameters for two
separate energy channels (0--4\,keV and 14--45\,keV; see Nowak et al.
\nocite{nowak:99a} 1999a) are also presented in Table~\ref{tab:qpo}, and
the fits are shown in Fig.~\ref{fig:cyg}.

The \cyg\ PSDs show subtle, but statistically significant, features. Broad
peaks are seen at $\approx 0.3$\,Hz, 2\,Hz, 10\,Hz, and 50\,Hz.  The nearly
flat portion of the PSD at $\aproxlt 1$\,Hz shows some sign of a peak.  We
use the same model that we fit to \gx, namely a zero frequency-centered
Lorentzian, a QPO and its harmonic, plus two additional QPO, and obtain
reasonable fits.  The reduced $\chi^2$ for these fits were significantly
lower ($\Delta \chi^2 > 150$ for eight extra degrees of freedom) than that
for the doublely broken power law fits discussed in Nowak et al.
\shortcite{nowak:99a}.  Note that the lowest frequency `QPO' is identified
with the portion of the PSD at $\aproxlt 1$\,Hz, and its harmonic is
identified as having both larger amplitude and greater $Q$ value, in
contrast to the fits to \gx.

Although these fits are a significant improvement over the doublely broken
power law fits of Nowak et al. \shortcite{nowak:99a}, there are large
residuals, especially near 0.2\,Hz.  We thus considered another fit
consisting of five QPO features plus one harmonic (or possibly
sub-harmonic, as the QPO associated with the higher of the two related
frequencies had a larger amplitude).  The results for these fits are
presented in Table~\ref{tab:cygqpo}, and the fits are also shown in
Fig.~\ref{fig:cyg}.  This model resulted in significant improvements to the
fits ($\Delta \chi^2 \approx 50$---90, for four extra degrees of freedom);
however, compared to the fits presented in Table~\ref{tab:qpo}, there was
relatively little change in the parameters for the highest frequency QPO
fit components.

\begin{table*}
\caption{\small Parameters for PSD Fits to \cyg\ Data. All frequencies are
  units of Hz.
  \protect{\label{tab:cygqpo}}} 
\center{
\begin{tabular}{lcccccccc}
\hline
\hline
\noalign{\vskip 3pt}
&& 0--4\,keV &&&& 14--45\,keV \\
\noalign{\vskip 3pt}
Component & $R$ & $f$ & $Q$ & $\chi^2$/DoF & $R$ & $f$ & $Q$ 
     & $\chi^2$/DoF \\
\noalign{\vskip 3pt}
\hline
\noalign{\vskip 3pt}
${\rm QPO_1}$ & \errtwo{0.29}{0.02}{0.02} & \errtwo{0.07}{0.02}{0.06} &
\errtwo{0.25}{0.12}{0.21} & 71.4/53 & \errtwo{0.22}{0.01}{0.01} &
\errtwo{0.10}{0.02}{0.03} & \errtwo{0.37}{0.11}{0.14} & 65.4/53 \\
\noalign{\vskip 3pt}
${\rm QPO_2}$ & \errtwo{0.04}{0.03}{0.01} & \errtwo{0.17}{0.01}{0.02} &
\errtwo{12.6}{51.3}{9.8} && \errtwo{0.03}{0.02}{0.02} &
\errtwo{0.16}{0.01}{0.02} & \errtwo{6.0}{24.6}{3.7} \\
\noalign{\vskip 3pt}
${\rm QPO_{sh}}$ & \errtwo{0.06}{0.16}{0.05} & \errtwo{0.82}{0.05}{0.05} &
\errtwo{2.3}{9.4}{1.9} && \errtwo{0.04}{0.16}{0.04} &
\errtwo{0.77}{0.10}{0.13} & \errtwo{1.9}{1.8}{1.7} \\
\noalign{\vskip 3pt}
${\rm QPO_3}$ & \errtwo{0.07}{0.02}{0.02} & --- & \errtwo{3.2}{2.5}{1.0} &&
\errtwo{0.15}{0.03}{0.02} & --- & \errtwo{1.2}{0.3}{0.3} \\
\noalign{\vskip 3pt}
${\rm QPO_4}$ & \errtwo{0.23}{0.04}{0.03} & \errtwo{0.45}{0.45}{0.40} &
\errtwo{0.16}{0.12}{0.14} && \errtwo{0.15}{0.02}{0.02} &
\errtwo{4.4}{1.8}{1.7} & \errtwo{0.60}{0.31}{0.24} \\
\noalign{\vskip 3pt}
${\rm QPO_5}$ & \errtwo{0.07}{0.02}{0.03} & \errtwo{13.9}{45.1}{12.5} &
\errtwo{0.21}{0.85}{0.19} && \errtwo{0.05}{0.02}{0.01} &
\errtwo{42.4}{6.3}{9.1} & \errtwo{1.9}{1.2}{0.9} \\
\noalign{\vskip 3pt}
\hline
\end{tabular}
}
\end{table*}

In these models, what was fit as a zero frequency-centered Lorentzian is
now fit as a strong, broad QPO plus a weaker, narrow QPO.  Whether these
fit components should be regarded as separate physical features, or whether
the improvement in the fit is merely an indication of the inadequacy of the
zero frequency-centered Lorentzian in describing the low frequency PSD is,
of course, unclear.  Also along these lines, we note that although we have
constrained the features at $\approx 0.8$\,Hz and 1.6\,Hz to be harmonics
of each other, the fits are slightly improved ($\Delta \chi^2=7.0$ for the
0--4\,keV band and $\Delta \chi^2=8.9$ for the 14--45\,keV band) if the
frequencies are allowed to float freely.  Note also that there are still
positive residuals at $\aproxgt 10^{-3}$\,Hz. The above points, as we
discuss further below, serve to highlight the fact that when the PSD
features are as broad and as subtle as they are in both \cyg\ and \gx\, the
intepretation of what is and is not to be considered a ``QPO'' becomes much
more ambiguous. This needs to be borne in mind in the discussion below of
the correlations between fit features.

\section{Discussion}\label{sec:discuss}

The evidence for a third, high-frequency feature in the composite PSD of
\gx\ is clear.  Removing this feature, the $\chi^2$ increases to $442$.
Here we should note that to some extent the fits are affected by the fact
that separate observations with slightly varying PSD properties have been
averaged together.  Some broadening of the ``QPO features'' is expected as
a result, and the $Q$ values for features fit to the summed PSD should be
considered as lower limits.  Likewise, the fitted frequencies and
amplitudes of the variability features should be viewed as indicative of an
average value, but not strictly applicable to any of the individual PSD
used in the combined observation.  Furthermore, their error bars should be
viewed as lower limits.  Along these lines, we note that although the
presence of a third feature is indicated, it is almost as well-modelled
($\chi^2=207$) by a zero frequency-centered Lorentzian ($\propto
[1+(f/f_b)^2]^{-1}$) with a break frequency of $f_b = 23\pm3$\,Hz.

The question arises as to what extent this third variability feature might
be an artifact of averaging together seven separate observations.  This is
of some concern; however, we note that in terms of well-measured
properties, the seven individual PSDs are very similar to each other.
Using the normalization of Belloni \& Hasinger \cite{belloni:90a}, at
frequencies $<10$\,Hz (where all seven individual PSDs have good
statistics) the variance of the noise subtracted PSDs is $\aproxlt 25\%$.
Specifically,
\begin{equation}
{P_s(f)}^{-1} {\left ( \sum_i [P_s(f) - P_i(f)]^2 \right )^{1/2}} \aproxlt 0.25
  ~~,
\end{equation}
at all frequencies $f \le 10$\,Hz, where $P_i(f)$ are the noise subtacted
PSD of the individual observations, and $P_s(f)$ is the noise subtracted
PSD of the summed observation.  The variance of the mean is lower by a
factor of $\sqrt{6}$ and therefore is $\aproxlt 10\%$. This result is
essentially unchanged if we weight the observations by count rate.  The
fact that some variation exists from observation to observation means that
the error bars that we used should be viewed as lower limits, and the
derived $\chi^2$ should be viewed as upper limits.  Also as regards
possible systematic variations, NWD showed the broad features at $\sim
0.35$ and $2.5$\,Hz have fitted frequencies that ranged only from $\approx
0.3$--$0.4$\,Hz and $2$--$3$\,Hz, respectively, among the individual
observations \emph{and} among separate energy bands within each of these
observations. Given the similarities among the PSDs, it would be unusual
for the strong third variability feature present in the combined
observation to be an artifact of the PSD averaging.  (We note that the
third variability feature discussed below for \cyg\ was for a single, short
observation.)

\begin{figure}
\resizebox{\hsize}{!}{\includegraphics{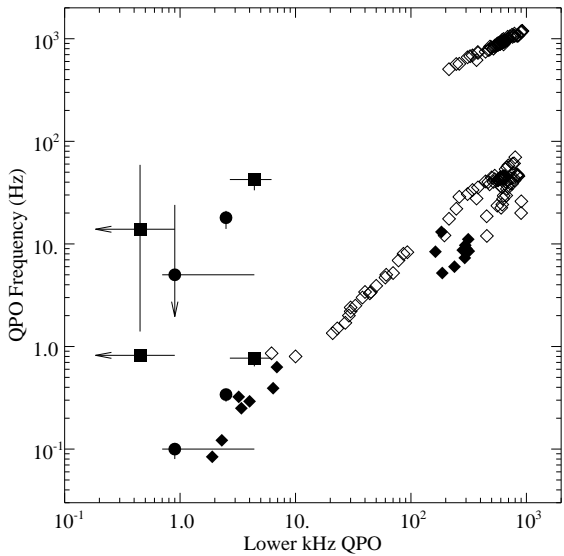}}
\caption{\small The QPO frequencies presented here (filled circles are for
  \gx, filled squares are for \cyg), with 90\% CL error bars, overlaid on
  the same graph as presented in Psaltis, Belloni, \& van der Klis
  \protect{\shortcite{psaltis:99a}}.  Open diamonds correspond to neutron
  star candidates, filled diamonds to black hole candidates (error bars are
  not shown for either).  Note: QPO identified as ``normal branch
  oscillations'' are also shown on this plot (see PBK).
  \protect{\label{fig:corr}}}
\end{figure}

Taking all three features in \gx\ as real and not an artifact of the PSD
summation, these features are relatively broad (even considering possible
systematic broadening due to averaging) compared to high frequency QPO seen,
for example, in neutron stars.  Statistically, they all have $Q>0$ at a
high significance level; however, they are all far from narrow, periodic
features.  The QPO at the higher frequency end of the correlation presented
by PBK all tend to be much narrower features (in terms of FWHM) than those
presented here.  Nevertheless, if we take these features and identify the
low-frequency QPO with the HBO, the middle-frequency QPO with the
lower-frequency kHz QPO, and the high-frequency QPO with the
upper-frequency kHz QPO, then they approximately fall along the
correlations suggested by PBK.  We show the above data overlaid on the
suggested correlation of PBK in Fig.~\ref{fig:corr}.  This is true for the
composite PSD as well for the PSD of the faintest observation from NWD,
keeping in mind that for this latter observation the error bars on the
highest frequency feature are for the local minimum and that the presence
of this latter feature is not strongly constrained. We also note that the
correlation between the frequency, $f_0$, of the zero frequency-centered
Lorentzian and the frequency, $f_1$, of the lowest frequency QPO is in the
same sense as the break frequency/QPO frequency correlation discussed by
Wijnands \& van der Klis \shortcite{wijnands:99a}.

The fact that the ``HBO'' and ``lower-frequency kHz QPO'' of \gx\ appear to
lie approximately along the suggested correlation was already noted by PBK.
What is new here is that there appears to be a third QPO that extends the
lower-frequency kHz QPO/upper-frequency kHz QPO correlation downward by
nearly two orders of magnitude in Fourier frequency. The high-frequency QPO
of \gx\ appears to lie slightly below the extrapolation of the
lower-frequency kHz QPO/upper-frequency kHz QPO correlation; however,
assuming the correlation to be $\propto f^\alpha$, a slope change of
$\delta \alpha \sim -0.2$ is all that is required between the low-frequency
\gx\ point and the high-frequency trend. Furthermore, some theories (e.g.,
the relativistic precession theory of Stella \& Vietri \nocite{stella:98a}
1998; Merloni et al. \nocite{merloni:99aa} 1999; Stella et al.
\nocite{stella:99a} 1999) suggest that one should not expect a single power
law over the whole range of the putative correlation. Specific predicted
frequency correlations for the relativistic precession theory can be found
in Stella et al. \shortcite{stella:99a}.

The features fit to the \cyg\ PSD are slightly more problematic to fit onto
this trend.  In light of the fact that the best fit consists of five QPO
features plus one harmonic, there are the questions of which fit components
are to be considered as ``true QPO'' (as opposed to merel a fit artifact)
and which are to be associated with which feature in the putative
correlation.  In Fig.~\ref{fig:corr}, we have taken the fits of
Table~\ref{tab:cygqpo} and associated ${\rm QPO_5}$ with the
upper-frequency kHz QPO, ${\rm QPO_4}$ with the lower-frequency kHz QPO,
and ${\rm QPO_{3}}$--- the sub-harmonic to the stronger 0.16\,Hz feature---
with the HBO.  With these associations, the features fit to the high energy
(14--45\,keV) PSD of \cyg\ agree with suggested trend, whereas the features
fit to the low energy (0--4\,keV) PSD of \cyg\ deviate from the trend,
mostly due to the extremely low frequency of the ``lower-frequency kHz
QPO''.  We note, however, that the frequency-break of this component occurs
at roughly the same location in the low and high energy PSDs.  The fact
that the feature in the low energy PSD has an extremely low $Q\approx0.2$
leads to a very low fit-frequency in order to yield a break in the same
location as the somewhat narrower feature ($Q\approx 0.6$) in the high
energy PSD.  If one performs a joint fit to the 0--4\,keV and 14--45\,keV
PSDs, constraining the QPO frequencies (but not the amplitudes or widths)
to be the same in both bands, a reasonable fit is obtained with $\chi^2 =
156$ for 111 degrees of freedom.  The resulting fit frequencies are
comparable to the frequencies fit to the 14--45\,keV band in
Table~\ref{tab:cygqpo}.

It is difficult to test the correlations suggested by PBK since most of the
$Q$ values for the fitted features are very low, making the features very
subtle.  Another, more serious problem, is the question of identification
of the features.  One alternative fit to the \cyg\ PSDs would be to place a
zero frequency-centered Lorentzian with break frequency $\approx 0.01$ to
remove the low frequency residuals seen in Fig.~\ref{fig:cyg}, and then to
identify ${\rm QPO_1}$ and ${\rm QPO_2}$ as the HBO and a harmonic, ${\rm
  QPO_3}$ and ${\rm QPO_h}$ as the lower-frequency kHz QPO and a harmonic,
${\rm QPO_4}$ as the upper-frequency kHz QPO, and ${\rm QPO_5}$ as a new,
``fourth QPO''.  [The theoretical existence of a fourth such frequency has
been suggested by Psaltis \& Norman (2000), for example.] These frequencies
would also fit on the suggested trend of PBK, except now ${\rm QPO_4}$ in
the low energy PBK would be too low a frequency to fall on the trend for
the upper-frequency kHz QPO. \emph{At present, there is no unambiguous,
  rigorous method of associating a fitted feature with the suggested trend
  of PBK.}

Even given these caveats regarding correlations between the various
fit-components, it is tempting to associate each fit-component of the PSD
with a ``resonance'' in the variability properties of the accretion flow
(see Psaltis \& Norman \nocite{psaltis:99b} 2000).  This is counter to the
models of Kazanas et al. \shortcite{kazanas:97a} and Poutanen \& Fabian
\shortcite{poutanen:99a}, for example, that essentially postulate a single
``response'' for the variability properties of the accreting system.  NWD
phenomenologically elaborated upon the concept of each PSD fit-component
corresponding to a separate physical mechanism by further postulating that
each of these system ``responses'' had a separate ``driver'', uncorrelated
with the variability-drivers of the other PSD components\footnote{Even if a
  system has a set of independent responses to a source of input
  fluctuation or noise, for example the accretion disk responses discussed
  by Psaltis \& Norman \shortcite{psaltis:99b}, the net outputs will be
  perfectly correlated (i.e., have a coherence function of unity) if they
  are responding to the same source of noise fluctuations.  This point is
  discussed in further detail by Bendat \& Piersol \shortcite{bendat},
  Vaughan \& Nowak \shortcite{vaughan:97a}, and NWD.}.  The net Fourier
frequency-dependent phase lags (or, equivalently, time lags$=$phase
lags/$2\pi f$) and coherence function [$\gamma^2(f)$, a measure of the
degree of \emph{linear} correlation between two lightcurves; see Bendat \&
Piersol \nocite{bendat} 1986, Vaughan \& Nowak \nocite{vaughan:97a} 1997]
between soft and hard variability were then given by a combination of the
phase lags and coherence functions for each individual fit-component of the
PSD (see NWD, \S4 and Fig.~8; Vaughan \& Nowak \nocite{vaughan:97a} 1997).

\begin{figure}
\resizebox{\hsize}{!}{\includegraphics{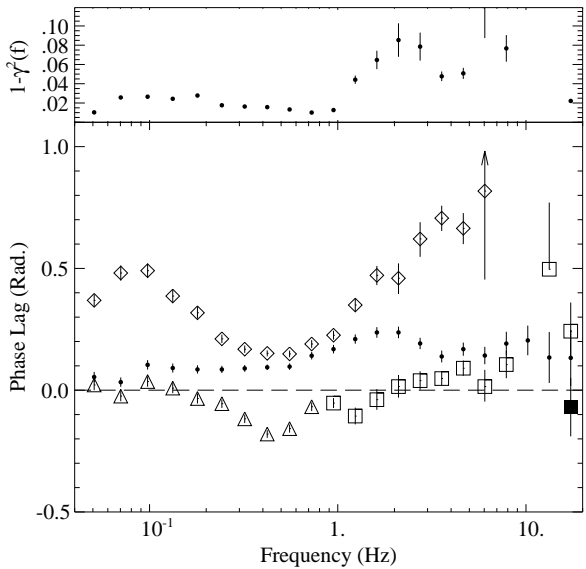}}
\caption{\small {\it Bottom Panel:} A possible decomposition of the net Fourier
  frequency-dependent phase lags between soft (0--3.9\,keV) and hard
  (10.8--21.9\,keV) X-ray variability (filled circles) into individual
  components associated with the PSD fit components shown in
  Fig.~\protect{\ref{fig:comppsd}}.  Triangles correspond to the zero
  frequency-centered Lorentzian, diamonds to the low-frequency QPO
  (including harmonic), clear squares to the middle-frequency QPO, and
  filled squares to the high-frequency QPO.  Positive values correspond to
  the hard X-ray variability lagging the soft. {\it Top Panel:} The
  coherence function, $\gamma^2(f)$, associated with the net phase lag.
  \protect{\label{fig:decomp}}}
\end{figure}

NWD noted that for the case of \gx\, wherever one PSD fit component
dominated (for example, near 0.3\,Hz in Fig.~\ref{fig:comppsd}, where the
low-frequency QPO dominates), the coherence function would be near unity
and there would be an approximately flat Fourier phase lag shelf when
comparing soft and hard variability. Wherever two PSD fit components
crossed one another, there would be a slight dip in the coherence function
and a transition from one characteristic Fourier phase lag to another (see
Fig.~\ref{fig:decomp}).  Fig.~\ref{fig:comppsd} shows that at any given
Fourier frequency, typically two PSD fit components dominate.  Ignoring the
other two fit components, at any given frequency we can then calculate the
phase lag between soft and hard X-ray variability for the remaining two PSD
fit components by simultaneously fitting the measured net phase lag and the
measured net coherence function\footnote{Note that there is still an
  ambiguity in the values of the Fourier phases, as there are two
  independent solutions for a two-component fit to the measured phase lag
  and coherence function.  Specifically, see Fig.~8 of NWD, where
  reflecting the phases of the individual PSD components through the net
  measured phase lag yields and equally valid solution.  In our solution
  that follows, at low Fourier frequency we choose the phase lag for the
  zero frequency-centered Lorentzian to be the value closest to the net
  measured phase lag.  Then, progressing to higher Fourier frequencies, we
  choose solutions that are most nearly continuous with the lower frequency
  values.}.

We have carried out such a procedure for the second brightest \gx\ 
observation (which showed the strongest QPO at $\approx 0.3$\,Hz) presented
in NWD.  We have fit the 0--21.9\,keV PSD of this observation with the same
type of model as discussed in Table~\ref{tab:qpo}. For this non-composite
PSD fit, constraints are again weakest for the ``third QPO'' fit-component.
Furthermore, we have \emph{assumed} this PSD shape and amplitude for both
the 0--3.9\,keV and 10.8--21.9\,keV PSDs.  (Fitting each energy band
individually, no constraints can be made on the ``third QPO''.)  We then
followed the procedures outlined in \S4 of NWD (see specifically eqs. 4 and
5) for fitting the phase lag and coherence function, except that instead of
fitting four constant (as a function of Fourier frequency) phases over the
entire data set, we fit two phases at any given Fourier frequency.
Specifically, at any given Fourier frequency, we fit the phase lags for the
two strongest PSD fit-components, \emph{assuming the other two PSD fit
  components to be negligible}.  Error bars do not account for
uncertainties in the PSD fit parameters, and only represent that portion of
the error due to the uncertainty in the measured phase lag and measured
coherence function.  The results for this decomposition (which, as we
discuss above, is \emph{not} unique) are presented in
Fig.~\ref{fig:decomp}.

Fig.~\ref{fig:decomp} is presented more for illustrative purposes rather
than as a completely rigorous and definitive fit.  The things to note about
this possible decomposition are the following.  If the QPO fit components
each represent a separate, independent physical process, then the
deviations from unity coherence can be understood within the context of
such a model as well.  At low frequency the net phase/time lags are most
characteristic of the zero frequency-centered Lorentzian, at middle
frequencies they are most representative of the low frequency QPO, and at
moderately high frequencies they are most representative of the middle
frequency QPO.  At $\approx 0.3$\,Hz, there is a recovery of the coherence
function to near unity values that would be associated with the low
frequency QPO becoming dominant in the PSD.  Although it is difficult to
accurately measure the high frequency phase/time lags and coherence
function, the fact that the middle and high frequency fit components to the
PSD have comparable amplitudes at $f\aproxgt 10$\,Hz could explain the
strong loss of coherence at these frequencies (which is characteristic of
Cyg~X-1 as well; see Nowak et al. \nocite{nowak:99a} 1999a).  Although the
net phase lag indicates that hard photons lag the soft photons, it is
possible for portions of the individual variability components to show
exactly the opposite behavior.

\section{Summary}\label{sec:summary}

The main results of this work can be summarized as follows.

\begin{itemize}
\item A composite power spectrum of \gx\ is well-fit by a model that
  consists of a zero frequency-centered Lorentzian, a moderate width
  ($Q\sim 1$--$2$) quasi-periodic oscillation and its harmonic, plus two
  additional, broad ($Q\sim0.5$--1) QPOs.
\item The frequencies of these three QPO components approximately agree
  with the correlations among horizontal branch
  oscillations/lower-frequency kHz QPOs/upper-frequency kHz QPOs suggested
  by Psaltis, Belloni, \& van der Klis \shortcite{psaltis:99a}.
\item Fits to the PSD of \cyg\ are more problematic. Multiple QPO
  components provide the best fits to the PSD as a function of observed
  energy band. Depending upon the identifications made, the high energy
  band PSD fits can be made consistent with the correlations suggested by
  Psaltis, Belloni, \& van der Klis \shortcite{psaltis:99a}. The low energy
  band PSD fits show deviations from these correlations, although a joint
  fit of the low- and high-energy PSD does allow for a single set of
  frequencies consistent with the suggested correlations.
\item One can construct a phenomenological model where the net Fourier
  frequency-dependent phase lag and coherence between soft and hard X-ray
  variability can be expressed as the combination of phase lags and
  coherences from the individual fit components to the PSD.
\item If such a model for the phase lags and coherences is correct, then
  although the net phase lag indicates that soft variability leads the hard
  variability, the trend may be exactly opposite for some of the individual
  variability components.  A \emph{requirement} of this model is that
  although the individual components have coherent variability between soft
  and hard photons, the variability components are incoherent with one
  another. This not only requires that each component be representative of
  a different ``response'' within the accretion flow, but also that each
  component has a separate variability ``driver''
\end{itemize}

The correlations between the different fit-components seems at least to be
very suggestive.  It is tempting to associate all of these quasi-periodic
features observed in both neutron star and black hole systems with a common
set of phenomena intrinsic to the accretion flow itself, and not to the
properties of the compact object such as the presence of a surface or a
magnetic field.  Such may not be the case in reality, however.  The
presence of what appear to be four distinct features (the zero
frequency-centered Lorentzian and three broad, quasi-periodic features) in
the PSD of \gx, coupled with suggestive evidence from the Fourier
frequency-dependent phase lags and coherence function (see also NWD), at
the least seems to be arguing for models with multiple, distinct sources of
variability (e.g., Psaltis \& Norman \nocite{psaltis:99b} 2000), and
against models with a single ``type'' of variability (e.g., Kazanas et al.
1997, Poutanen \& Fabian 1999\nocite{kazanas:97a,poutanen:99a}).

\smallskip This research has been supported by NASA grants NAG5-3225 and
NAG5-4731. Dimitrios Psaltis kindly provided the data for
Fig.~\ref{fig:corr} and valuable comments.  I also would like to thank
Andrew Hamilton for useful discussions, the hospitality of the Aspen Center
for Physics, the participants of the ``X-ray Probes of Relativistic
Astrophysics'' workshop for many useful and stimulating discussions, and
the hospitality of P. Coppi, C. Bailyn, and the Yale Astronomy department
while this work was being completed.


\end{document}